# Zinc Oxide-Based Piezoelectric Pressure Sensor

Victor K. Samoei and Ahalapitiya H. Jayatissa*

Nanotechnology and MEMS Laboratory, Department of Mechanical, Industrial and Manufacturing Engineering (MIME), MS 312, University of Toledo, Toledo, OH 43606, USA.
**\*Correspondence:** ahalapitiya.jayatissa@utoledo.edu; Tel: +1(419)-530-8245

**Abstract:** This paper reports the application of zinc oxide (ZnO) in the pressure sensors that can be integrated with a microelectromechanical system (MEMS). ZnO is one of the materials that has received a great deal of attention due to its unique properties of being a semiconductor with wide bandgap and piezoelectric effects. The simpler crystal growth mechanisms of ZnO have resulted in a lower cost of ZnO-based sensors. Different types of pressure sensors based on ZnO sensing elements have also been explored. A thin circular ZnO film was simulated as a piezoelectric sensor employing the finite element method in COMSOL. The pressure applied on the thin film surface was varied and a boundary point probe was used to study the displacement field and voltage at the center of the membrane. The displacement field and voltage induced by pressure vary linearly with increasing pressure on the ZnO layer. Also. the method used in this paper was applied to different piezoelectric materials, such as barium titanate ($BaTiO_3$), polyvinylidene fluoride (PVDF), and gallium arsenide (GaAs) that were studied by other groups, and similar conclusions were made. These simulations can be used in the design of piezoelectric sensors and the optimization of the sensitivity and performance of the materials used in pressure sensor applications.

## 1. Introduction

Pressure sensors are essential components in many systems, they are electronic devices used to detect the pressure state of an element, these devices can be used both in commercial and industrial applications. The applications of pressure sensors are widely growing due to the technological advancement that has led to a wide range of demands. Pressure sensors are mainly based on materials and physical properties such as piezoelectric, capacitive, and piezoresistive among others. These devices are determined by the change in the design, technology, cost, performance, and applications [1]. These factors have led to the rapid development of MEMS highly sensitive, and high-performance pressure sensors, the rapid development and the wide application of sensors have led to the new development of different types of pressure sensors such as flexible pressure sensors that can be applied in robotics, biomedical, and other electronic devices [2].

The mathematical expression of pressure is force per unit area, therefore, when pressure is applied to the sensor, the element in the sensor, such as a diaphragm changes shape resulting in mechanical movement, which produces an electrical signal that can be detected and measured by electrical devices [2]. A good sensor should be highly sensitive to the property intended to be measured with a high variation on the output compared to small variations on the magnitude applied. The sensor should not deform permanently when the intended measured property is applied, also sensors should be sensitive, accurate, and have a high-resolution ratio. Pressure sensors have been applied in measuring different kinds of measurements such as speed, altitude, fluids, and gas flow [3]. Therefore, it is important to develop sensors that are cost-effective, scalable, and light [3]. The MEMS pressure sensors currently dominate the market due to the increased miniaturization and performance of these devices. The MEMS devices have gained wide applications in automotive, medical electronics, aerospace, and many other disciplines [4, 5].

This paper will report on current methods used in the fabrication of ZnO-based pressure sensors and the fabrication of sensors with new emerging technologies in the application of pressure sensors. ZnO pressure sensors have grown widely due to their vast application in the field of engineering. The properties of ZnO including piezoelectric effect, non-toxicity, inexpensiveness, and responsiveness to UV have led to a wide application on pressure sensors, solar cells, and transistors on the pressure sensor [5]. The material used in pressure sensors mainly depends on the application of the sensor, however, ZnO has proven to be the material of choice compared to other materials because few materials are piezoelectric, semiconductors, and have a wide bandgap. The properties of ZnO are discussed that have led to the wide application of pressure sensors is [5].

### 1.1. Classification of pressure sensors

Pressure sensors can be classified in many ways: they can be classified according to the targeted type of pressure measurement, application, or sensing elements used such as piezoresistive, piezoelectric, and capacitive pressure sensors [6].

A piezoelectric sensor is a device that uses the piezoelectric effect to detect applied mechanical stress to measure the change in pressure, strain, or force and convert it into an electronic signal. This unique behavior is called the piezoelectric effect where, certain materials generate an electric charge when they are subjected to mechanical stress [7]. This effect exists naturally in a few crystals such as quartz, berlinite, nitrate barium titanate, single-crystal leads, and zirconate titanate (PZT) [5,7,8]. The influence of mechanical stress causes shifting of the positive and negative charge centers in materials resulting in an external electric field that either compresses or stretches the piezoelectric material. Fig. 1(a) illustrates this phenomenon that induces electric polarization when stress is applied [7,8,9]. The ionic crystal such as quartz in a neutral state is not polarized but when pressure is applied on the surface of the crystal, oxygen and silicon ions are displaced generating electric dipoles. Similarly, Fig.1(b) shows the mechanism of piezoelectric behavior of ZnO, which has a piezoelectric polarization along c-axis when subjected to strain or stress [10].

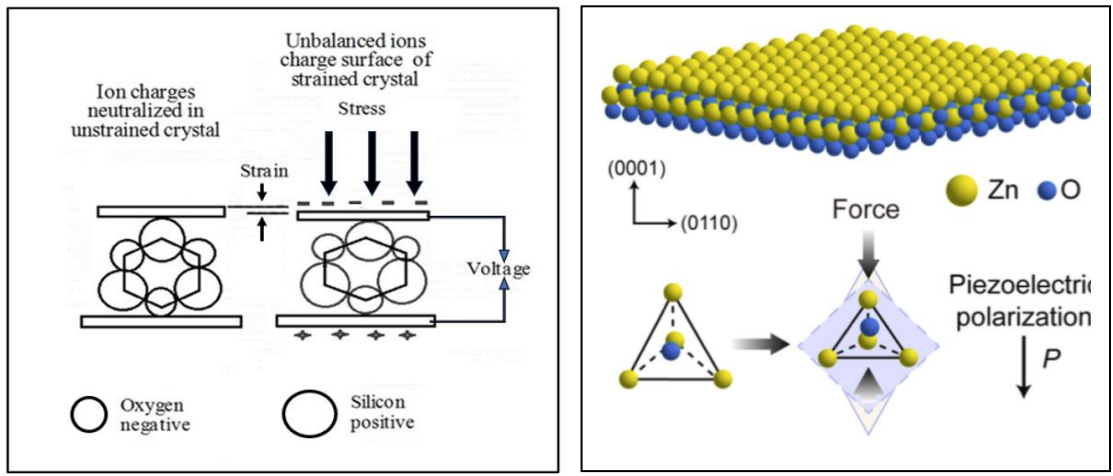

**Fig. 1**: Diagram indicating the charge deflection on the poles when a strain is applied on the surface of the material [8,10] (with permission).

The electric field $E$ generated when a potential difference is applied on the surface can be calculated using the eq. (1). The 3D constitutive equations of piezoelectric material are given by eq. (2) and (3), where, eq. (2) represents the stress to electric potential also known as the direct effect and eq. (3) represent the converse effect that is from electric stimulus to strain [9]. The matrix form of the constitutive equations is given by Eq. (4).

$$D = \varepsilon_r E = \varepsilon_0 E(1 + x) = \varepsilon_0 E = P, \tag{1}$$

$$D_m = d_m + E_{m_k}^T E_k, \tag{2}$$

$$S_i = S_{ij}^E T_j + d_{m_i} E_m, \tag{3}$$

$$\begin{bmatrix} S \\ D \end{bmatrix} = \begin{bmatrix} s^E & d^t \\ d & \varepsilon^T \end{bmatrix} \begin{bmatrix} T \\ E \end{bmatrix}. \tag{4}$$

Where, $S_i$ is the strain, $T_i$ is the stress, $D_i$ the electric displacement, $E_k$ is the applied field, $s$ is the elastic compliance, $d$ is piezoelectric coupling, and $\varepsilon$ are permittivity matrices. $\varepsilon_r$ is the relative permittivity and $\varepsilon_0$ the permittivity of free space, the superscript $t$ stands for the transpose, and the superscripts $E$ and $T$ denote that the respective constants are evaluated at constant electric field and stress [9]. The piezoelectric parameters and measurements for BaTiO$_3$, ZnO, PZT, and PVDF are given in Table 1. Here, $d$ and $k$ are a piezoelectric charge and the coupling factor respectively in their respective directions.

**Table 1**: Properties of piezoelectric materials: BaTiO$_3$, ZnO, PZT, and PVDF.

| Material Chemical formula | T°C | d$_{33}$ | d$_{31}$ | Coupling factor $k$ | k$_{15}$ | Relative permittivity ($\varepsilon / \varepsilon_0$) | Ref. |
|---|---|---|---|---|---|---|---|
| BaTiO$_3$ | 115 | 190 | -78 | 0.21 | 0.48 | 0.21 | [7] |
| ZnO | Room temp | 5.9 | -5.0 | 0.33 | | 8.2 | [11] |
| PZT | Room temp | 60-130 | -120 | 0.57-0.69 | | | [11] |
| PVDF | Room temp | 30 | -20 | 0.11 | | 12 | [12] |

The following Table 2 gives a summary of different forms of the piezoelectric constitutive equations with the piezoelectric constants: *d, e, g,* and *h* defined in Table 2 [8]. Piezoelectric has a high modulus of elasticity compared to other materials and for this reason, the piezoelectric pressure sensors are highly preferred because it has a high natural frequency, high voltage, high charge sensitivity and it can produce a high output when little stress is applied.

**Table 2**: Summary of the piezoelectric constitutive equations [8].

| Piezoelectric equations | Definitions of the constants | SI Units |
|---|---|---|
| $D = d\sigma + \varepsilon(d)^\sigma E$ | $d$= charge density/applied stress | C/N |
| $D = e\varepsilon + \varepsilon(d)^\varepsilon E$ | $e$=charge density/applied strain | C/m |
| $E = -g\sigma_{+D}/_{\varepsilon(d)^\sigma}$ | $g$=field/applied strain | v/m/N |
| $E = -h\varepsilon^{+D}/_\varepsilon (d)^\varepsilon$ | $h$=field/applied strain | V/m |
| $\varepsilon = dE + S^E\sigma$ | $d$=strain/applied field | m/V |
| $\varepsilon = gD + S^D\sigma$ | $g$=strain/applied charge density | m/C |
| $\sigma = -eE + EY^E\varepsilon$ | $e$=stress/applied field | N/V/m |
| $\sigma = -hD + E(Y)^D\varepsilon$ | $h$=stress/applied charge density | N/C |

### 1.2. *Piezoresistive pressure sensor*

These pressure sensors measure pressure by detecting the change in electrical resistance in a material when mechanical stress that causes deformation in a material is applied [14]. The four-strain sensors are arranged in a pattern that they coincide with the area of maximum sensitivity for external pressure on the thin diaphragm that induces deflection. The following Eq. (5) is used to measure the resistance of the piezo resistor.

$$R = \frac{(\rho L)}{(A)}, \tag{5}$$

where, $\rho$, $l$, and $A$ are resistivity, length, and area of the piezo resistor. This pressure sensor is widely used due to its low cost, good sensitivity and it is relatively simple to construct [15] . They are commonly used because of the possibility of batch production and linearity [16]. This pressure sensor has a high gauge factor, which limits the operating temperature [17]. The capacitive pressure sensors are advantageous over piezoresistive sensors because they use less power, no resistor mounted on the sensor diaphragm is needed, and are applicable in many mechanical systems because of their high-pressure sensitivity, low-temperature drift, and they are easy to scale down [13].

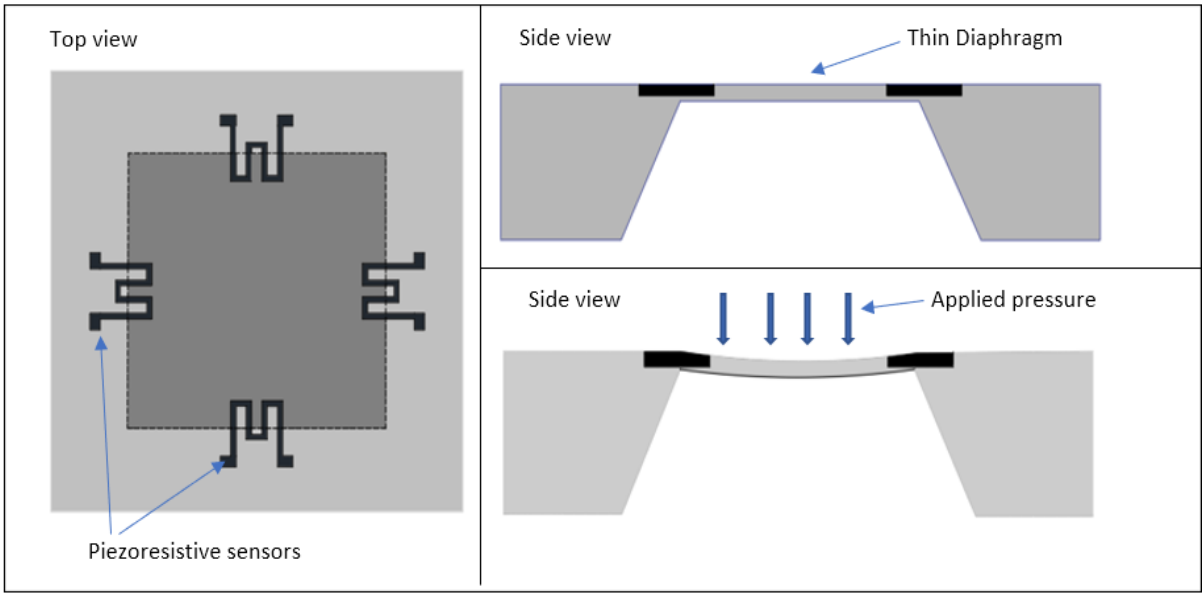

**Fig. 2**. Piezoresistive-based pressure sensor with a thin silicon plate and strain sensors placed on the edge of the thin silicon plate.

### 1.3. Governing equations for diaphragm membrane deflections.

The common geometries used for the pressure sensors are square, circular, or rectangular these geometries are designed to measure maximum deflection when pressure is applied to the diaphragm. Typically, the small deflection and thin plate theory equations are applied for micromachined pressure sensors. The thin plate concept refers to a condition where, the thickness $h$ of the plate is approximately one-tenth of the radius ($\frac{a}{10}$). The following theory of plate eq. (6) is used to obtain the maximum deflection ($w$) of a circular thin plate that is fixed at the ends with uniform pressure applied perpendicular to the surface, From eq. (6), the deflection $w$ is directly proportional to the applied pressure $P$ [5,12,16,18].

$$w(r) = \frac{Pa^4}{64D}\left[1 - \left(\frac{r}{a}\right)^2\right]^2. \tag{6}$$

Where, $r$ is the radial coordinate, $a$ is the radius of the diaphragm, $P$ is the differential pressure and $D$ is the flexural rigidity. The flexural rigidity $D$ can be calculated by the following equation.

$$D = \frac{Eh^3}{12(1-v^2)}. \tag{7}$$

Where, $E$, $h$, and $v$ are Young's modulus of the material used, the thickness of the diaphragm, and the Poisson's ratio, respectively [14]. The maximum deflection for square and rectangular plates is given by the following eq. (8) [5]. In this case, $a$ and $b$ are the lengths of the diaphragm and $\alpha$ is the numerical factor depending on the ratio of $a$ and $b$. These equations can be applied to optimize the design and fabrications of the sensors.

$$\begin{cases} w = \alpha \frac{pa^3}{D} & \text{for } a < b \\ w = \alpha \frac{pa^3}{D} & \text{for } a < b \end{cases}. \tag{8}$$

1                                         where, $\alpha = 1.26*10^{-3} *12(1-v^2)$.

## 2.0. Material Selection for MEMS-based Pressure Sensors

The applications of the pressure sensors are the main factors that determine the material to be used in the sensing element of the pressure sensor. The micromachined pressure sensors are advancing rapidly and fabrication of low to highly sensitive sensors is growing. With this growth, materials that can replace old metallic diaphragms are considered for the fabrication of the sensors. Materials that are low cost, lightweight, non-toxic, bio-friendly, and have high reliability are important for the current pressure sensors. For example, polymer materials have gained wide applications on pressure sensors because of their properties such as flexibility, structural stability, good thermal and electrical properties [16]. Materials such as stainless, titanium, quartz, silicon, and sapphire are used to design the capacitative diaphragms for better sensitivity [5]. Fabrication of micro pressure sensors applies mainly to silicon and its compounds since material properties and fabrication processes such as surface and bulk micromachining are well defined for silicon [17].

The nano-polysilicon is another material that has been used to fabricate highly sensitive pressure sensors. This material is deposited by a low-pressure chemical vapor deposition process on a thin film to the high resistivity wafer. The output voltage against the pressure of the nano-polysilicon is a linear graph with an upward trend. The external pressure acting normal to the diaphragm with different thicknesses is used to analyze the sensitivity. The sensitivity is inversely proportional to the thickness of the silicon diaphragm [23]. The channel resistors of a thin-film transducer can be used to improve the sensitivity in the micro pressure sensors. In the recent development of pressure sensors, ZnO has been used to record the pressure by utilizing ZnO piezoelectric properties. ZnO has a large piezoelectric constant compared to other materials such as Gallium nitride, GaN that has similar properties to ZnO is expensive for both scaling and deposition. It also requires high deposition temperatures. Cadmium could be used to achieve similar pressure sensor properties but it is toxic and expensive [26].

## 2.1 Properties of zinc oxide

The ZnO has three crystal structures; wurtzite, zinc-blende, and rock-salt, which is rarely noticed. Wurtzite is the naturally occurring structure of ZnO and its unit cells are hexagonal with two lattice parameters. It is the most common and stable in environmental conditions [20, 21].

The ZnO is a piezoelectric semiconductor with a wide bandgap with an energy gap of 3.37 eV at room temperature [26] and a high piezoelectric coefficient that gives high sensitivity [3]. it can be integrated into pressure sensors display because of its amenability to the low-temperature range, wet chemical etching and is stable at high temperature (1800 ºC) [3, 21]. ZnO is most suitable for the fabrication of microdevices because it is easily etched in acids and alkalis [27]. Other properties of ZnO that are attractive to its diverse applications in sensors are the high refractive index (1.95 - 2.10) and the high specific surface area of the active grades [28]. Table 3 gives the summary of the basic properties of ZnO that makes it most suitable for numerous application including veterinary science and antibacterial [28].

**Table 3:** The relative advantages and disadvantages of the three sensors discussed [19].

| Type | Advantage | Disadvantage |
|---|---|---|
| Piezoelectric | Dynamic response<br>High bandwidth<br>High natural frequency,<br>High voltage<br>High charge sensitivity | Temperature sensitivity |
| Piezoresistive | Simple to construct<br>Low cost<br>Good sensitivity | Stiff<br>Nonlinear<br>Hysteresis<br>Temperature sensitivity |
| Capacitive | Low cost<br>Good sensitivity<br>Available for commercial A/D chips | Complex electronics<br>hysteresis |

**Table 4:** The summary of the basic properties of ZnO [26].

|  | Properties | Description |
|---|---|---|
| 1 | Appearance | White solid |
| 2 | Molecular weight | 18.38 g/mol |
| 3 | Crystal structure | Wurtzite |
| 4 | Coordinate geometry | Tetrahedral |
| 5 | Band Gap | 3.37 eV |
| 6 | Solubility in water | 0.0004 % (17.8 ºC) |
| 7 | Refractive index ($\mu_o$) | 2.0041 |
| 8 | Density | 5.606 g/cm$^3$ |
| 9 | Melting Point | 1975 ºC |
| 10 | Flash Point | 1436 ºC |

The other properties of ZnO that need to be discussed concerning pressure sensors are mechanical properties, this involves the discussion of the piezoelectric, bulk moduli, hardness, and yield strength [27]. The five elastic constants for the hexagonal crystal are $C_{11}$ $C_{12}$ $C_{33}$ $C_{13}$ and $C_{44}$, where, $C_{11}$ and $C_{33}$ are longitudinal modes in the direction [1000 and 0001]. The bulk modulus can be related to elastic constants by the following equation.

$$B = \frac{(C_{11}+C_{12})C_{33}-2C_{13}^2}{C_{11}+C_{12}+2C_{33}-4C_{13}}. \tag{9}$$

The piezoelectricity of ZnO is determined by the relation between the charges and the electric field [30]. The equation that can be used for ZnO in determining the electromechanical process can be written as:

$$D_m = d_{m_i} + eE. \qquad (10)$$

Where, $E$ is the applied electric field, $D$ is the piezo strain matrix with units of $\frac{m}{v}$, $d$ is the stress tensor, $e$ is the permittivity of measured constant stress, $D$ is the electric displacement. The dielectric constant of reactivity sputtered ZnO is 13.4. The value of stress that can be developed inside the piezoelectric material is found to be 12.4 pc/N on the $d_{33}$ plane. Young's and shear modulus $E$ and $G$ can be found using the relation of bulk modulus $B$ and Poisson's ratio $v$ relation expressed in the equations:

$$E = 3B(1 - 2v) \text{ and } G = \frac{E}{2(1+v)}. \qquad (11)$$

## 2.2. Fabrication of ZnO film for pressure sensors

Many techniques have been studied for the synthesis of ZnO thin films. The few of these techniques are rf magnetron sputtering [31], ion plating [32], sol-gel process (chemical vapor deposition) [32], molecular beam epitaxy [34], spin coating [34] and pulse laser deposition [36]. The sol-gel process is also known as the wet chemical deposition method is preferred and most promising because it is easy to handle, low cost, and easy to control the final product also this technique gives rise to high-quality films and that applies to large scale production [38].

Bulk growth of ZnO is undertaken in three methods; hydrothermal, vapor phase, and melt growth. These processes are difficult to be carried out because of the high pressure needed, which makes it difficult to control the process. This process is preferred in obtaining large crystals of ZnO since it is almost impossible to use the sputtering technique. The hydrothermal method is normally carried out because it has been well established hence it is suitable to process [37,38].

In the rf magnetron sputtering, a high purity ZnO target is used to grow ZnO on a substrate using a magnetron sputter system. Inert gases are used in the sputtering process because they do not participate in film growth. ZnO can also be grown using pure Zinc target, argon, and oxygen gas mixtures passed at different ratios. Oxygen can react with the Zn-target to form ZnO. This technique has great potential to attain a great high depository rate, enhanced adhesion, great control on the film thickness, and maintain uniformity [27,31,39]. This process is usually preferred because it is simple, needs low temperature for carrying out the process and low cost compared to other techniques. When rf power is increased the deposition rate increases, the power enhances the bombardment of the target with electrons and ions from plasma leading to increased sputtering [37,40,41]. The deposition process is affected by the distance between the substrate and the target, the deposition rate decreases with the increase in the distance, also the electrons get on the surface with low energy affecting the film growth dynamics at the surface of the substrate [27,28]. The gas pressure inside the chamber affects the sputtering rate also. The deposition rate tends to increase with the increase in the chamber pressure. The chamber pressure is always at the range of 2 mT and 10 mT for rf sputtering [42-44].

## 3.0. Simulation and Modeling of Piezoelectric Pressure Sensors

Structural designs have been modeled and simulated as piezoelectric devices using COMSOL Multiphysics software to analyze the parameters related to piezoelectric materials [45]. The pressure is applied on the surface of the material and the deflection in the $z$-direction is measured, the measured voltage and displacement varied linearly with increasing pressure. Analytical solution for small membrane with uniform applied pressure is given by the eq. (6). The analytical solution for the pressure sensor was found to be qualitatively similar to the results from COMSOL simulation when pressure was applied to the diaphragm [46].

### 3.1. Zinc oxide simulation as a piezoelectric sensor

The properties of ZnO were utilized to simulate a circular diaphragm that acts as a pressure sensor. As discussed, ZnO is a piezoelectric material, therefore, electric polarity is generated when mechanical stress is applied, and a converse effect occurs whereby mechanical stress is produced when a potential difference is applied [46]. COMSOL Multiphysics software is utilized for the finite element analysis of the geometry with a thin film circular diaphragm modeled for the study. ZnO is used as the material of choice to carry the simulation and utilize the piezoelectric properties. The material parameters of ZnO used for simulation are relative permittivity, density, Poisson's ratio, Young's modulus, elasticity matrix (ordering: $xx$, $yy$, $zz$, $yz$, $xz$, $xy$), compliance matrix (ordering: $xx$, $yy$, $zz$, $yz$, $xz$, $xy$) and coupling matrix. A boundary point probe is placed at the center of the circular diaphragm, the probe is used to read the voltage and deflection values at the center of the diaphragm when pressure is applied. Center location is chosen because that is the point that experiences the maximum deflection when pressure is applied to the diaphragm.

### 3.2. Simulation Results

The displacement and electrical potential values were obtained using the boundary point probe when varying pressure is applied on the surface of the diaphragm. Fig. 3 shows the stress plot of ZnO when pressure is applied on the surface (-$z$-direction). The electric displacement varied linearly with pressure as indicated in Fig. 4. The electric potential (V) against pressure is given in Fig. 5, which depicts similar characteristics as electric displacement and pressure. The data obtained for the electric displacement field, z-component (C/m$^2$) and electric potential were graphed against pressure applied. The corresponding plots are given in Fig. 4 and Fig. 5.

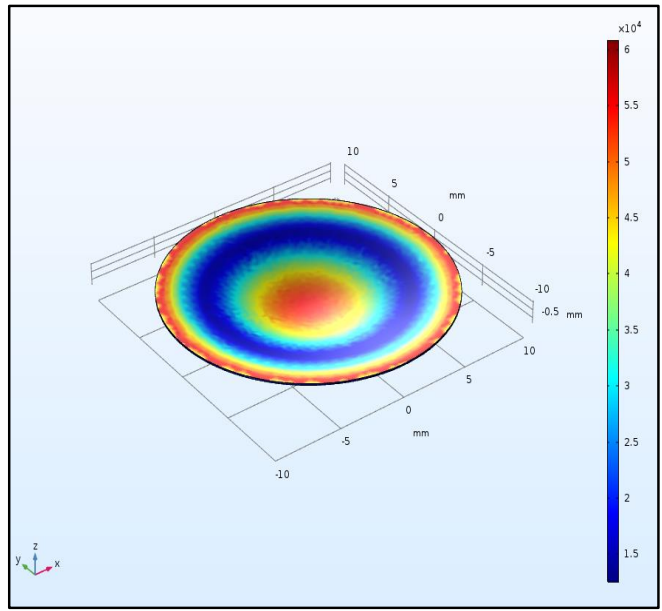

**Fig. 3.** von Mises stress (N/m$^2$) plot deformation of ZnO.

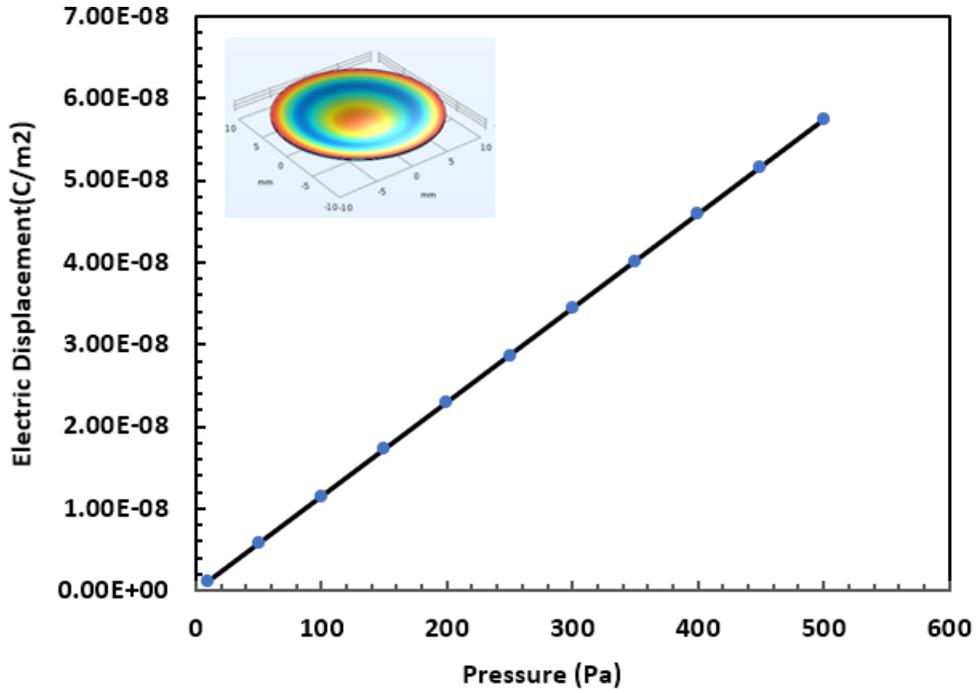

**Fig. 4:** Graph of the electric displacement field, z-component (C/m²) against pressure applied on the z-direction on the circular ZnO diaphragm.

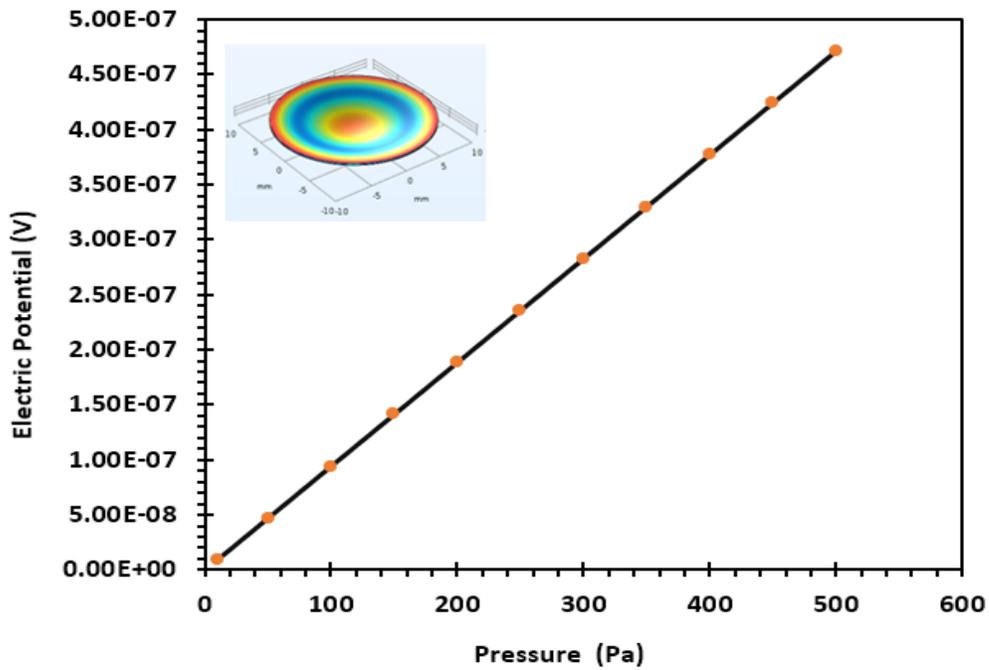

**Fig. 5:** The electric potential field, z-component (C/m²) against pressure applied on the z-direction on the circular ZnO diaphragm.

### 3.3. Comparison of MEMS piezoelectric materials for modeling of piezoelectric pressure sensors

A comparison of different MEMS piezoelectric materials that can be used on pressure sensors is considered in this section. The MEMS-based pressure sensors that have high sensitivity can be used in many areas such as automobile, aerospace, and biomedical to convert mechanical energy into electrical energy using MEMS technology as MEMS-based pressure sensors. First, three different MEMS materials were considered for comparison. The comparison is done by utilizing COMSOL Multiphysics software for the finite element analysis. The materials for the comparison are BaTiO$_3$ PVDF, and GaAs [48]. The structure utilized for the simulation is a square with rectangular sections at the edges. The three piezoelectric materials are applied at the most stressed regions of the structure used for the analysis as indicated in Fig. 6.

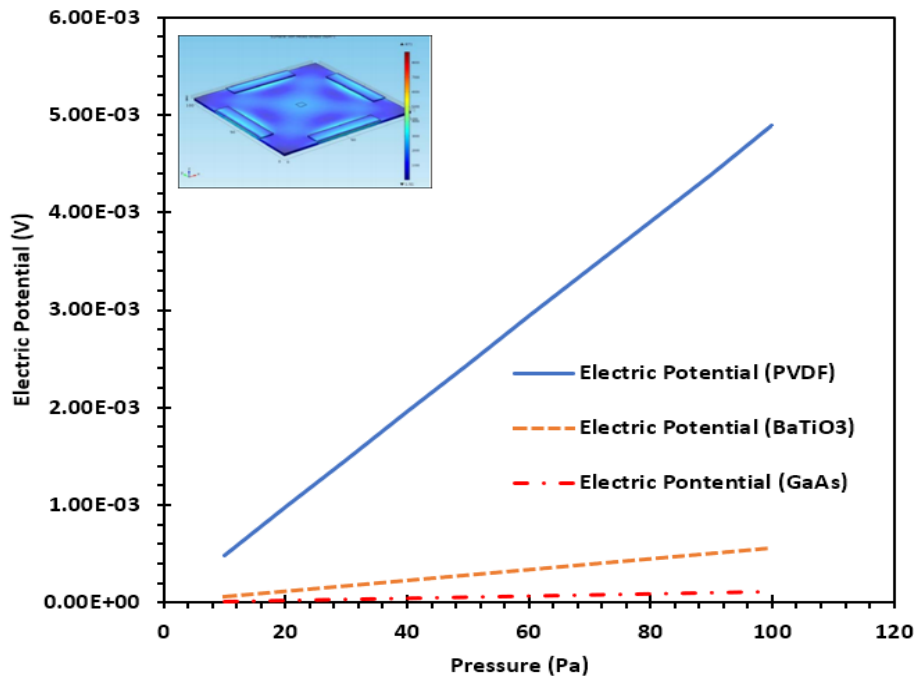

**Fig. 6.** The electrical potential against pressure for BaTiO$_3$, PVDF, and GaAs [47].

The materials are then analyzed separately by applying different boundary conditions. The pressure versus electrical potential plot is then determined for these materials. Pressure varies linearly with electrical potential as shown in Fig. 6. PVDF had greater variation indicating that it was a better piezoelectric material compared to the BaTiO$_3$ and GaAs. The same materials were then analyzed by utilizing a circular structure that was previously used for ZnO analysis in Fig. 3. The pressure is varied from 10 Pa to 100 Pa as shown in Fig. 7 and the corresponding values of electrical displacement (V) are shown in the y-axis. The variation of pressure and electrical displacement is linear for these materials. Both graphs indicated that PVDF exhibits higher piezoelectricity in comparison to the other two materials, which can be concluded that PVDF produces high voltage to small variation of pressure and hence Fig. 6 and Fig. 7. A similar comparison is then carried out for ZnS and BaTiO$_3$ piezoelectric materials. Simulation parameters of piezoelectric materials are done using the COMSOL Multiphysics software. Here, the circular diaphragm in Fig. 3 that was previously studied using ZnO material is utilized again for the comparison, the structure has finely meshed, and a stationary type of study is selected for simulation results. The deformation of the structure used is achieved on the Z-direction when uniform pressure is applied on the surface. Maximum displacement is achieved at the center of the structure and minimum displacement is at the fixed ends as shown in Fig. 3. Using boundary probes, a linear variation of electrical potential (V) and electrical displacement (mm) against different values of pressure is obtained as shown in Fig. 8 and Fig. 9. ZnS is highly piezoresistive and highly sensitive compared to BaTiO$_3$ due to its high piezoelectric coefficient. A comparative simulation study of these materials has been done using a cantilever-based MEMS pressure sensor and the sensitivity results obtained for electric displacement and electric potential are comparable both indicating a linear variation with pressure [49]. The simulation results for different piezoelectric materials can be used as an overview for the optimization of performance for different piezoelectric MEMS-based pressure sensors.

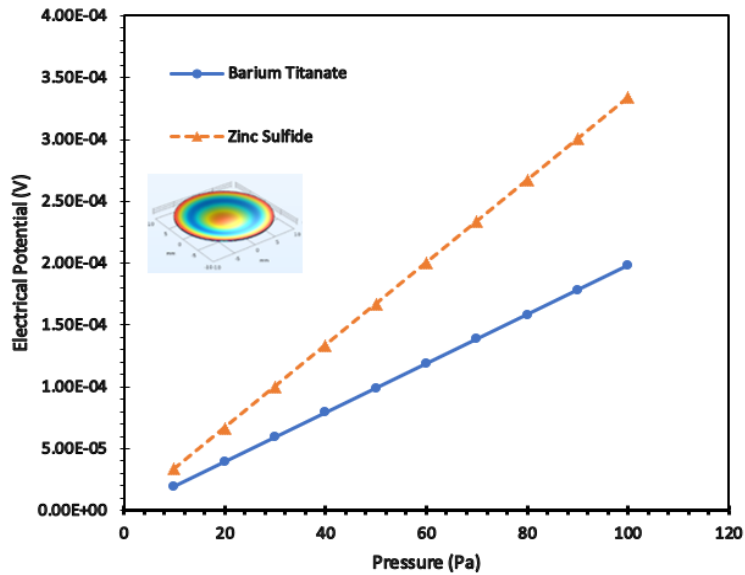

**Fig. 7.** The electric potential (V) against pressure (Pa) for PVDF, GaAs, and BaTiO$_3$ was obtained by using a circular diaphragm in Fig. 3.

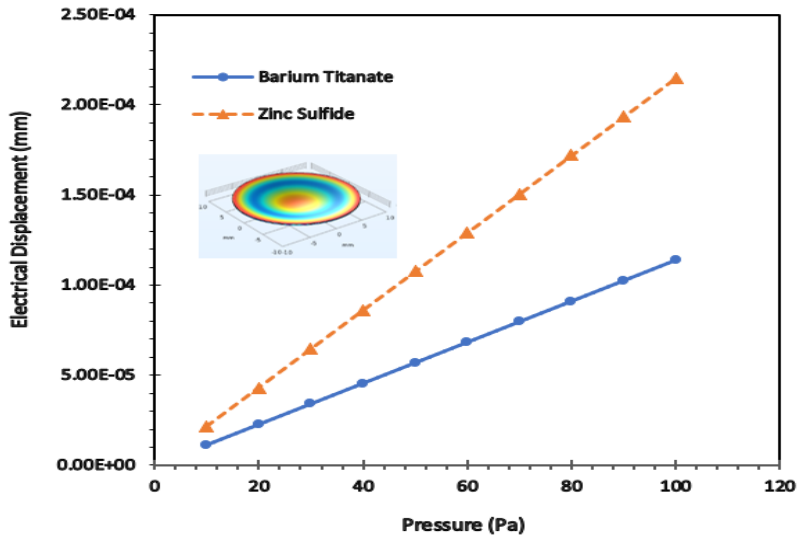

**Fig. 8.** The linear variation of electrical displacement (mm) against increasing pressure (Pa) for BaTiO$_3$ and ZnS.

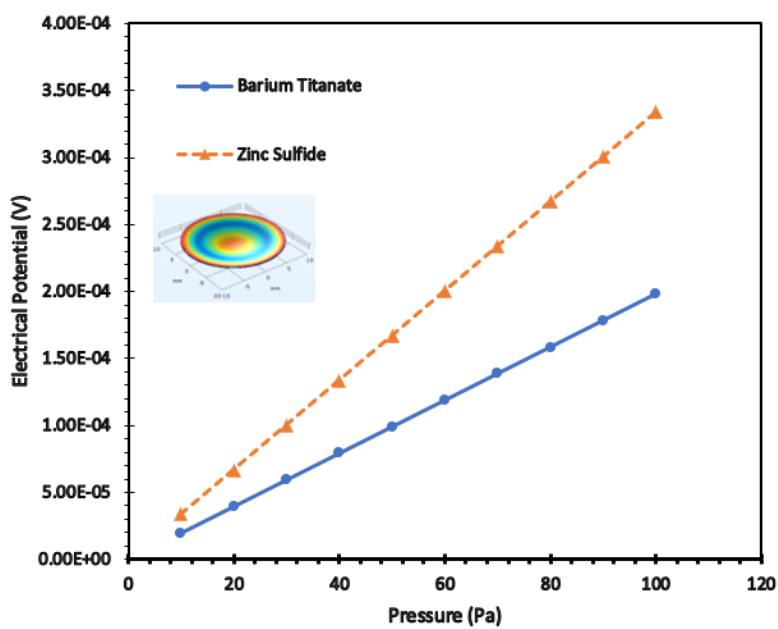

**Fig. 9.** The linear variation of electric potential (V) against increasing pressure (N/m$^2$) for BaTiO$_3$ and ZnS.

### 3.4. Advantages and Disadvantages of ZnO as a Piezoelectric pressure sensor material

Several reasons can be attributed to the wide application of ZnO material on pressure sensors. First, ZnO is a naturally occurring n-type semiconductor, therefore, the conductivity is generally high since the effective mass of the electrons is lower compared to p-type semiconductor also the n-type semiconductor is easier to dope compared to p-type that are extremely difficult to carry the doping process. The natural occurrence of the ZnO makes it abundantly available making it the cheapest material for designing pressure sensors. Another advantage of using ZnO on pressure sensors is that it has been widely studied and methods of obtaining ZnO nanoparticles are well explained. ZnO is eco-friendly and biocompatible with the human body, which has led to the current application of implantable pressure sensors. The other advantages of ZnO that have led to the wide application of pressure sensors in the medical field are its non-toxic nature, low thermal expansion, high thermal capacity, and high melting point. Another reason is that no link indicates that ZnO is carcinogenic, genotoxic, and reproduction toxic. ZnO being a piezoelectric material gives electric potential when pressure is applied, however, the generated output voltage is low. The low output voltage from these pressure sensors requires an amplifier, which makes the overall setup bulky.

### 4.0. Conclusion

This paper presents the applications of ZnO on pressure sensors. As technology advances, more complex and sophisticated devices are required in response to the newly emerging needs in the market. The miniaturization of these devices has led to advancements in MEMS sensing technology. ZnO has been applied in many sensing technologies and has recently gained more interest due to its physical and chemical properties. The production of ZnO has risen and the techniques of production have evolved rapidly due to large-scale demand by the industries. The method of production is determined by the properties required for the various applications, process throughput, and the cost of production. Overall, there are new emerging applications of ZnO in many areas including electronics, solar cells, liquid crystal displays and, in the sensor, and actuators. The piezoelectric properties of ZnO make it suitable for various sensor applications as discussed. ZnO and other piezoelectric materials have been studied with the finite element analysis of the structure and design of the pressure sensors. It is clearly shown that the material displacement and electrical potential have a linearly increasing trend to the applied pressure. The study shows that the prospects of MEMS technology and its applications of ZnO in pressure sensors look bright as the demand is rapidly growing.